\documentclass[aps,prd,twocolumn,floatfix, preprintnumbers,nofootinbib]{revtex4}
\usepackage[pdftex]{graphicx}
\usepackage[mathscr]{eucal}
\usepackage[pdftex]{hyperref}
\usepackage{amsmath,amssymb,bm,yfonts}
\usepackage[utf8]{inputenc}

\newif\ifarxiv
\arxivtrue

\begin		{document}

\def\w{\Omega}

\title
    {
    Hairy black resonators and the AdS$_4$ superradiant instability
    }

\author{Paul~M.~Chesler}

\date{\today}

\begin{abstract}
The superradiant instability of Kerr-AdS black holes is studied by 
numerically solving the full 3+1 dimensional Einstein equations. 
We find that the superradiant instability results in a two stage process with distinct initial and secondary instabilities.
At the end of the secondary instability the geometry oscillates at several distinct fundamental 
frequencies --- a multi-oscillating black hole.  The multi-oscillating black hole is remarkably close to a black resonator, albeit with a bit of gravitational hair.
During the hairy black resonator epoch, the evolution of the horizon area is consistent with the exponential approach to a constant.
By employing different seed perturbations in the initial Kerr-AdS geometry, we also demonstrate that the  black resonator's hair is not
unique.  In the dual quantum field theory description, rotation invariance is spontaneously broken
and the energy density is negative in some regions, signaling an exotic state of matter which does not relax to 
a stationary configuration.
\end{abstract}


\maketitle
\parskip	2pt plus 1pt minus 1pt

\section{Introduction}

Waves with suitably tuned frequency $\omega$ and azimuthal quantum number $m$ 
scattering off rotating objects can be amplified via superradiance \cite{zeldovich,Starobinsky:1973aij}. 
Of particular interest are rotating black holes, 
which contain ergoregions from which it is possible to extract energy and angular momentum \cite{Penrose:1969pc,Penrose:1971uk}. 
Half a century ago it was pointed out that if waves can be confined with something akin to a mirror, so outgoing waves are reflected back inwards, then
repeated interactions with a black hole can lead to exponential
growth of the wave amplitude, potentially converting a significant fraction of the black hole's 
mass into radiation \cite{Press:1972zz}.  
The dynamics of superradiant instabilities --- also known as black hole bombs --- and their final state
are of considerable of interest to a variety of fields including early universe cosmology \cite{Pani:2013hpa}, particle physics 
and gravitational wave astrophysics \cite{Arvanitaki:2016qwi,East:2017ovw,East:2018glu,Siemonsen:2019ebd},
astrophysical jets \cite{Brito:2015oca}, and phase transitions in holographic duality \cite{Hartnoll:2011fn,Chesler:2014gya}.  
For a detailed review of superradiance and black hole bombs see Ref.~\cite{Brito:2015oca}.

Perhaps the purest manifestation of a black hole bomb
is that of the AdS$_4$ superradiant instability.  
The dynamics of the system are governed by the 3+1 dimensional vacuum Einstein's equations 
with a negative cosmological constant and asymptotically global AdS boundary conditions.  
The geometry contains a time-like boundary (conformally equivalent to sphere), which serves as a mirror from which gravitational waves are reflected inwards.
Via holographic duality  \cite{Maldacena:1997re}, this 
system has a dual interpretation as a 2+1 dimensional strongly coupled quantum field theory living on the boundary.

For two decades it has been known that sufficiently rapidly spinning Kerr-AdS black holes 
are susceptible to superradiant instabilities \cite{Hawking:1999dp,Cardoso:2004hs}.
The spectrum of unstable modes in the Kerr-AdS spacetime was studied in \cite{Cardoso:2013pza}.
It was subsequently shown that any black hole in asymptotically 
AdS spacetime with an ergoregion --- meaning a region where a Killing vector becomes spacelike ---
must be  unstable \cite{Green:2015kur}.  It has been suggested 
that the AdS$_4$ instability may have no end-state and can result in violations of cosmic censorship \cite{Niehoff:2015oga} .
Despite the long history of the AdS$_4$ superradiant instability,  
the question of what the final state is has thus far remained elusive.  
Our goal in this paper is to study this long standing problem with numerical relativity simulations.

Suppose the dominant unstable mode in the Kerr-AdS spacetime has (complex) frequency $\omega$ and 
azimuthal quantum number $m$.  At a linear level the mode's time and azimuthal angle dependence 
is then given by the exponential $e^{-i \omega t + i m \varphi}$.
Hence the associated gravitational wave rotates in $\varphi$ at angular velocity $ {\rm Re} \frac{\omega}{m}$ 
while slowly growing in amplitude like $e^{{\rm Im} \, \omega \, t}$.  If the superradiant instability was a one stage process,
with this mode and its harmonics merely plateauing after some time, the final state geometry would have a single Killing vector,
\begin{equation}
\label{eq:ResonatorKilling}
\mathcal K = \partial_t + \Omega \, \partial_\varphi,
\end{equation}
with $\Omega \approx  {\rm Re} \frac{\omega}{m}$.
Black holes with single Killing vector --- coined black resonators ---  were first constructed in Ref.~\cite{Dias:2015rxy}.   Their geometry rigidly 
rotates with angular velocity $\Omega$, meaning they oscillate 
with a single fundamental frequency, 
and they are thermodynamically preferred over the Kerr-AdS solution with the same mass and angular momentum.
However, they are also unstable \cite{Green:2015kur}.  Their angular velocities satisfy
\begin{equation}
\label{eq:resonatorinstablity}
\Omega > \frac{1}{L},
\end{equation}
which means $\mathcal K$ is always spacelike near the AdS boundary, signaling the presence of an ergoregion.
It is therefore reasonable to expect that additional instabilities occur at some stage of the evolution.
A natural guess is that the superradiant instability results in the Kerr-AdS geometry transitioning to a black resonator, which then experiences its own
distinct superradiant instabilities, thereby transitioning to another state.

Previously we numerically simulated the AdS$_4$ superradiant instability
and found evolution consistent with the transition of the Kerr-AdS geometry to an approximate black resonator geometry \cite{Chesler:2018txn}.
While secondary instabilities were also observed, numerical evolution was not carried out
long enough to ascertain their end state, leaving the final fate of the system uncertain.
Nevertheless, Ref.~\cite{Chesler:2018txn} found several unstable modes during the black resonator epoch. 
These modes oscillated at different fundamental frequencies and rotated with different angular velocities.
This suggests that the final state has no continuous symmetries.
In other words, it is reasonable to expect that the final state of the superradiant instability is a 
black hole geometry oscillating with several fundamental frequencies, and
without a Killing vector or ergoregion.  
Following the nomenclature of Ref.~\cite{Choptuik:2019zji}, 
we will refer to black holes oscillating with multiple fundamental frequencies as 
{multi-oscillating black holes}.
Multi-oscillating black hole solutions to an Einstein-scalar system in AdS$_5$ were recently studied in Ref.~\cite{Ishii:2021xmn}.

One of the primary challenges in numerically simulating 
superradiant instabilities is the slow growth rates associated with unstable modes.
Compounding this, the AdS$_4$ superradiant instability has no symmetries to exploit, meaning one must numerically solve the full 3+1 dimensional Einstein equations.
This is a common feature of superradiant instabilities with real fields, and consequently there are few 
simulations which reach the final state \cite{East:2018glu}.
In contrast, with complex fields it is possible to study superradiant instabilities in cylindrical symmetry \cite{East:2017ovw} or 
charged superradiance with spherical  symmetry \cite{Sanchis-Gual:2015lje,Bosch:2016vcp,Bosch:2019anc}, thereby allowing much faster simulations.

To study the AdS$_4$ superradiant instability one must employ fast and stable numerical algorithms.  
To this end, we use a characteristic evolution scheme we previously developed for asymptotically AdS spacetimes.
This scheme --- which has seen use in a wide variety of problems
\cite{Chesler:2008hg,Chesler:2009cy,Chesler:2010bi,Heller:2012km,vanderSchee:2012qj,
Adams:2013vsa,Heller:2013oxa,Casalderrey-Solana:2013aba,vanderSchee:2013pia,
Casalderrey-Solana:2013sxa,Arnold:2014jva,Chesler:2015wra,Chesler:2015fpa,
vanderSchee:2015rta,Chesler:2018txn,Ecker:2021ukv} and is reviewed in detail in Ref.~\cite{Chesler:2013lia} ---
is remarkably efficient, allowing even five dimensional problems to be solved on a single CPU \cite{Chesler:2015wra}.
This efficiency is largely due to the fact that within the scheme, the apparent horizon can naturally be chosen to lie at some \textit{fixed} radial coordinate.
That is, the entire computation domain exterior to the apparent horizon and interior to the AdS boundary can be chosen to be a static spherical shell.
This means the angular and radial directions can be discretized with tensor product grids, allowing efficient numerical  integration and differentiation operations.
Additionally, with tensor product grids it is easy to employ spectral and pseudo-spectral discretizations, which converge much faster than 
finite difference schemes \cite{boyd01}.  This allows far coarser grids to be used and ameliorates CFL instabilities, meaning one can also 
use a larger time step compared to a finite difference scheme.
We also judiciously choose our initial conditions so that the instabilities are reasonably fast, but that the Kerr-AdS black hole is not 
too close to extremality.  We do this because near-extremal black holes can develop structure near the horizon, requiring finer grids there
and longer run times.  For reference, each of our simulations runs in approximately five weeks on a 2020 MacBook Air.

We simulate the Kerr-AdS superradiant instability with two sets of seed perturbations, both with the same mass and spin.
In both cases we find that the AdS$_4$ superradiant instability results in a two stage process with distinct initial and secondary instabilities.
At the end of the secondary instability the geometry is that of a multi-oscillating black hole with several distinct fundamental frequencies.
The multi-oscillating black hole is remarkably close to a black resonator geometry, albeit with a bit of gravitational hair
localized far from the horizon.  We see no obvious signs of additional instabilities in the multi-oscillating epoch. 
Indeed, at late times the apparent horizon area is consistent with the exponential approach to a constant.
Multi-oscillating black holes are therefore a plausible candidate for the endpoint of the superradiant instability. 
We also demonstrate that hairy black resonators are not unique and depend on initial conditions.

An outline of our paper is as follows.  In Sec.~\ref{sec:Setup}
we outline the setup of our problem, including initial conditions and our numerical evolution scheme.
In Sec.~\ref{sec:results} we presents the results of our numerical simulations.  Finally,
in Sec.~\ref{sec:discussion} we discuss our results.

\section{Setup}
\label{sec:Setup}

We numerically solve the vacuum Einstein's equations with negative cosmological constant $\Lambda = -3/L^2$.  We set the AdS radius $L$ to unity.  
Our characteristic evolution scheme is reviewed in detail in Ref.~\cite{Chesler:2013lia}.
Here we outline the details salient for characteristic evolution with spherical coordinates and an appropriate choice of basis functions.

Our metric ansatz takes the form,
\vspace{-0.15cm}
\begin{equation}
\label{eq:metric}
ds^2 = \lambda^2 g_{\mu \nu}(x^\alpha,\lambda) dx^\mu dx^\nu + 2 dv d\lambda,
\end{equation}
with Greek indices $(\mu,\nu)$ running over the AdS boundary spacetime coordinates 
$x^\mu = \{v,\theta,\varphi\}$,
where $v$ is time and $\theta$ and $\varphi$ are the usual polar and azimuthal angles in spherical coordinates.  
The coordinate $\lambda$ is the AdS radial coordinate, with the AdS boundary located at $\lambda = \infty$.  Note lines of constant $x^\mu$ are infalling radial null geodesics affinely parameterized by $\lambda$.  Correspondingly, the metric ansatz (\ref{eq:metric}) is invariant under shifts in $\lambda$,
\begin{equation}
\label{eq:gaugetrans}
\lambda \to \lambda + \xi(x^\mu),
\end{equation}
for any function $\xi(x^\mu)$.  We exploit this residual diffeomorphism invariance to fix the location 
of the apparent horizon to be at $\lambda = 1$.  This means horizon excision is 
performed by restricting the computational domain to $\lambda \geq 1$.

Near the AdS boundary Einstein's equations can be solved with the power series expansion,
\begin{equation}
\label{eq:expansion}
g_{\mu \nu}(x^\alpha,\lambda) = g_{\mu \nu}^{(0)}(x^\alpha) + \dots +g_{\mu \nu}^{(3)}(x^\alpha)/\lambda^3 + O(1/\lambda^4).
\end{equation}
The expansion coefficient $g_{\mu \nu}^{(0)}$ is the AdS boundary metric (i.e. the metric the dual quantum field theory lives in).  As a boundary condition we fix 
\begin{equation}
\label{eq:boundarycondition}
g_{\mu \nu}^{(0)} = \eta_{\mu \nu},
\end{equation}
where
\begin{equation}
\eta_{\mu \nu} = {\rm diag}(-1,1,\sin^2 \theta),
\end{equation}
is the metric on the unit sphere.  
This boundary condition means gravitational waves are reflected off the AdS boundary.

A convenient diffeomorphism invariant observable is the stress tensor $T_{\mu \nu}$ in the 
dual quantum field theory, which is determined by $g_{\mu \nu}^{(3)}$ via \cite{deHaro:2000vlm},
\begin{equation}
T_{\mu \nu} = g_{\mu \nu}^{(3)} + \textstyle \frac{1}{3}\eta_{\mu \nu} g_{0 0}^{(3)}.
\end{equation}
Note that Einstein's equations imply $T_{\mu \nu}$ is traceless, $\eta^{\mu \nu} T_{\mu \nu} = 0$, 
and covariantly conserved,
\begin{equation}
\nabla_\mu T^{\mu \nu} = 0,
\end{equation}
where $\nabla_\mu$ is the covariant derivative under the boundary metric $\eta_{\mu \nu}$.  

Within our characteristic evolution scheme, evolution variables consist of the conserved 
densities $T^{0 \mu}$, a gauge parameter $\xi$ used to shift the horizon to be at $\lambda = 1$ via Eq.~(\ref{eq:gaugetrans}), and the rescaled angular metric,
\begin{equation}
\hat g_{ab} \equiv \sqrt{\frac{{\det \eta_{cd}}}{{\det g_{cd}}}}   g_{ab},
\end{equation}
with lower case latin indices running over the angular directions $\{\theta,\varphi \}$.
All other components of the metric are determined by non-dynamical equations, which are solved 
on each slice of constant $v$.  The rescaled metric satisfies  
\begin{equation}
\label{eq:deteq}
\det \hat g_{ab} = \det \eta_{ab} = \sin^2 \theta,
\end{equation}
which means $\hat g_{ab}$ contains two independent degrees of freedom.  
These two degrees of freedom encode the two propagating degrees of freedom in 
gravitational waves.  In our numerical simulations we decompose $\hat g_{ab}$ as follows, 
\begin{equation}
\label{eq:angulardecomp}
\hat g_{ab} = \left  [1  + \textstyle \frac{1}{2} h^{ab} h_{ab} \right ]^{1/2} \eta_{ab}  + h_{ab},
\end{equation}
where indices are raised with $\eta^{ab}$ and $h_{ab}$ is traceless, meaning $h^{a}_{\ a} = \eta^{ab} h_{ab} = 0$.  Note the 
expansion (\ref{eq:expansion}) and boundary condition (\ref{eq:boundarycondition}) imply that near the boundary $h_{ab} \sim 1/\lambda^3$.

\subsection{Discretization}

We expand the angular dependence of all functions in a basis of scalar, vector and tensor harmonics.  
These are eigenfunctions of the covariant Laplacian on the unit sphere. The scalar eigenfunctions are just spherical harmonics $y^{\ell m}$.%
\footnote{
Note we employ the convention $y^{\ell{-}m} = y^{\ell m*}$.}
  There are two vector harmonics, $\mathcal V_a^{s \ell m}$, which read
\cite{oldref}
\begin{subequations}
\label{eq:vectorharmonics}
	\begin{eqnarray}
	\mathcal V_a^{1\ell m} &=& {\textstyle \frac{1}{\sqrt{\ell (\ell + 1)}}} \nabla_a y^{\ell m}, \\
	\mathcal V_a^{2\ell m} &=&  {\textstyle \frac{1}{\sqrt{\ell (\ell + 1)}}} \epsilon_a^{\ b} \nabla_b y^{\ell m},
	\end{eqnarray}
\end{subequations}
where $\epsilon_a^{\ b}$ has non-zero components $\epsilon_\theta^{\ \varphi} = \csc \theta$ and $\epsilon_{\varphi}^{\ \theta} = -\sin \theta$.
$\mathcal V_a^{1\ell m}$ is longitudinal, meaning it points in the direction of 
$\nabla_a$, and $\mathcal V_a^{2\ell m}$ is transverse, meaning $\nabla^a\mathcal V_a^{2\ell m} = 0$ .  There are three symmetric tensor harmonics $\mathcal T_{ab}^{s \ell m}$.
However, since $h_{ab}$ is traceless, we only need the two traceless tensor harmonics, which read 
\cite{oldref}
\begin{subequations}
	\begin{eqnarray}
	\mathcal T_{ab}^{1\ell m} &=& {\textstyle \frac{1}{ \sqrt{\ell \left (\ell + 1)( \frac{\ell(\ell + 1)}{2} - 1 \right)}}} 
	\epsilon_{(a}^{\ \  c} \nabla_{b)} \nabla_c   y^{\ell m},
	\\
	\mathcal T^{2 \ell m}_{ab} &=& {\textstyle \frac{1}{\sqrt{\ell (\ell + 1)\left ( \frac{\ell(\ell + 1)}{2} - 1 \right)}}} [ \nabla_a \nabla_b
	+ {\textstyle\frac{\ell (\ell + 1)}{2}} \eta_{ab} ]  y^{\ell m}. \ \ \ \  \ \ \ 
	\end{eqnarray}
\end{subequations}
  
The scalar, vector and tensor harmonics are orthonormal and complete.  Upon expanding SO(3) scalar, vector, and tensor components of the metric and boundary stress in terms of these basis functions, angular derivatives can be computed by differentiating 
the basis functions themselves.  In order to efficiently transform between real space and mode space, we employ a Gauss-Legendre grid in $\theta$ with $\ell_{\rm max} + 1$ points and a Fourier grid in $\varphi$ with $2 \ell_{\rm max} + 1$ points.  For details see Ref.~\cite{boyd01}.  The grid points $\theta_i$ are the zeros of the $\ell_{\rm max} + 1$ order Legendre polynomial, 
\begin{equation}
P_{\ell_{\rm max}+1}(\cos \theta_i) = 0,
\end{equation}
and the associated integration weights read
\begin{equation}
w^\theta_i = \frac{2 }{(d P_{\ell_{\rm max}+1}/d\theta)^2}\bigg |_{\theta = \theta_i}.
\end{equation}
Likewise, the $\varphi$ grid points are
\begin{align}
&\varphi_j = \frac{2 \pi j}{2 \ell_{\rm max} + 1}, &&  j = 0,1,\dots, 2 \ell_{\rm max},&
\end{align}
with associated integration weights $w^\varphi_j = \frac{2 \pi}{2 \ell_{\rm max}+1}$.
The transformation between real space and mode space can then be performed using Gaussian quadrature.%
\footnote{Note that integration over $\varphi$ can also be performed via Fast Fourier Transforms.  However, with the 
small values of $\ell_{\rm max}$ we employ in this paper, we have found Gaussian quadrature to be faster.}
Using Gaussian quadrature, the orthonormality of the scalar, vector and tensor harmonics is exact up to angular momentum $\ell = \ell_{\rm max}$.  We choose $\ell_{\rm max} = 40$.

We note that the mode expansions contain approximately half as many degrees of freedom as that which live on the real space grid.  Because of this, we choose to use mode amplitudes as our evolution variables.  When computing nonlinear products, we simply transform the mode amplitudes to real space first.
 
For the radial dependence of the metric, we employ  an inverse radial coordinate $u \equiv \frac{1}{\lambda} \in (0,1)$ and expand the $u$ dependence in a pseudo-spectral basis 
of Chebyshev polynomials.
We employ domain decomposition in the $u$ direction with seven domains.  
The first domain interface lies at $u = 0.05$
and the remaining interfaces are equally spaced between $u = 0.05$ and $u = 1$.  We use seven points in the domain closest to the boundary and 14 points in all other domains.  Note that using fewer points in the domain closest to the boundary helps ameliorate CFL instabilities, thereby allowing larger time steps. 

We evolve forward in time $3060$ units using a $4^{\rm th}$ order Runge–Kutta method with constant time step $dv = 0.008$.  
To test convergence of our numerics we have also ran simulations with approximately 15\% more grid points in each direction, and with 20\% smaller time step.  Likewise, we also increased the filter parameters $\ell_0$ and $\sigma_0$ (discussed below) by approximately 15\%.  Reassuringly, these simulations, which were ran until time $v = 1500$, which encompasses the times during which the most structure exists, produced results indistinguishable from those presented in this paper.  
In particular, the differences in Fig.~\ref{fig:modeamps45} were far smaller than the widths of each line.

\begin{figure}[h!]
	\begin{center}
		\includegraphics[trim= 0 0 0 0,clip,scale=0.45]{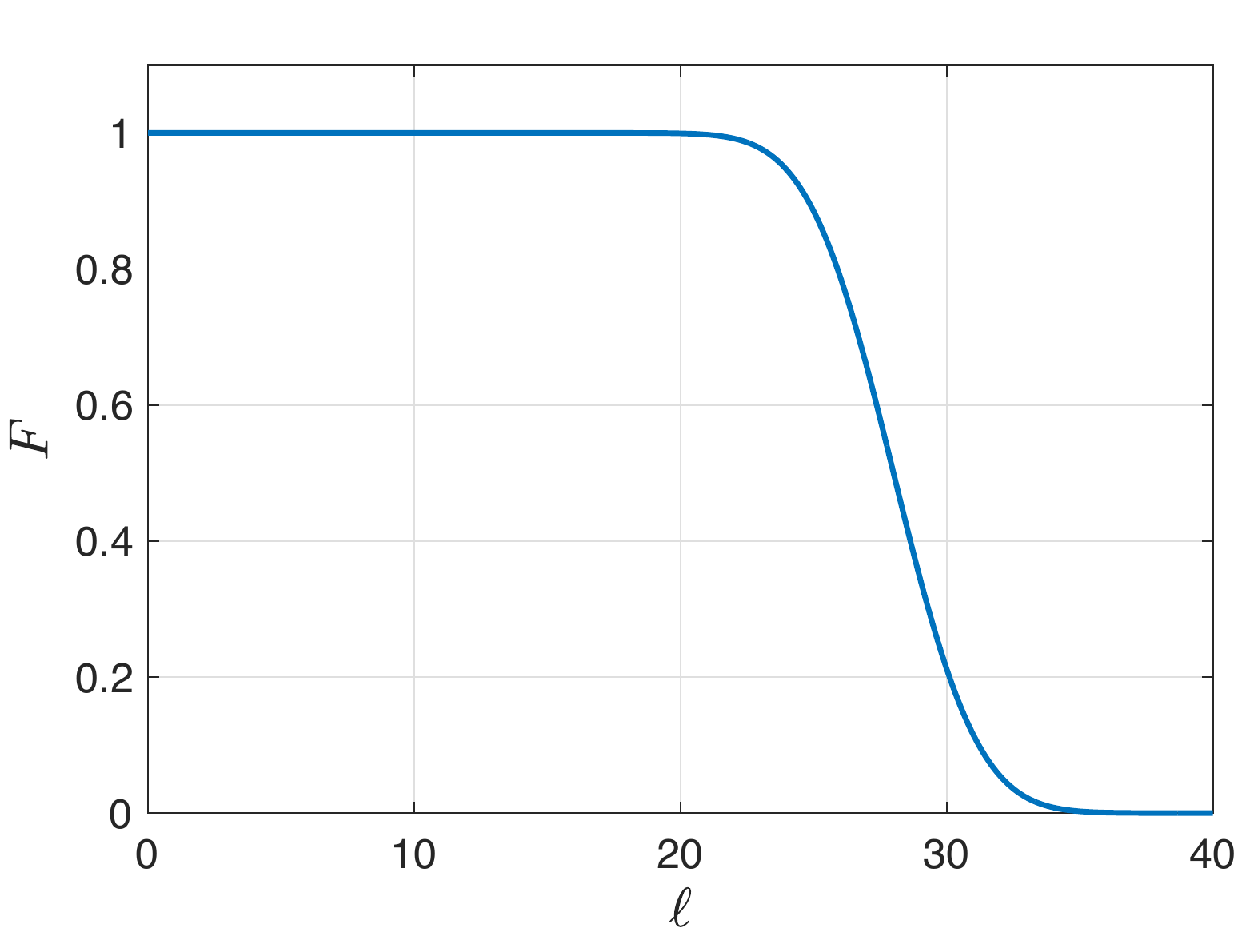}
		\caption{The angular filter function, Eq.~(\ref{eq:angularfilter}).  Modes with $\ell \lesssim 20$ are not appreciably modified by the filter.
		}
		\label{fig:filter}
	\end{center}
\end{figure}

\subsection{Filtering}

An important practical matter is filtering short wavelength excitations artificially generated during numerical evolution.
We apply a short wavelength filter directly to the time derivatives of the fields.  Following \cite{Chesler:2013lia}, to filter the radial direction, we simply interpolate the real space radial grid 
to a courser grid, and then reinterpolate back to the original grid.  For the course grid we use the same domain interface locations, but just with fewer points.  Specifically, we use one fewer point in the domain closest to the boundary, and two fewer points in all other domains.

To filter in the angular directions, at each time step we multiply the time derivatives of the mode amplitudes by the filter function
\begin{equation}
\label{eq:angularfilter}
F(\ell) \equiv \frac{1}{2} \left [1 + \rm{erf}\left (- \textstyle \frac{(\ell-\ell_0)}{\sqrt{2} \sigma} \right) \right ].
\end{equation}
where
\begin{align}
&\ell_0 = 28, &&\sigma = 2.5.&
\end{align}
This function, plotted in Fig.~\ref{fig:filter}, is ostensibly a regularized step function.  Note that at $\ell = 15$ we have $1-F \approx 10^{-7}$ and at $\ell = 20$ we have $1-F \approx 7 \times 10^{-4}$.  Hence this filter does not appreciably modify low angular momentum modes (e.g. those with $\ell \lesssim 20$).  

\begin{table*}[t!]
\begin{tabular}{@{\extracolsep{5pt}}c|cccc}
IC & $\alpha_2$ & $\alpha_3$ & $\alpha_4$  
\\
\hline
\hline
A &  $2.745 \times 10^{-2}$ \hspace{6mm} & $-6.863\times 10^{-3}$\hspace{9mm}  & $0$ & 
\\[2pt]
B  & $5.490 \times 10^{-2} e^{i \pi/3}$ & $-1.373\times 10^{-2} e^{-i\pi/3}$ & $3.768\times 10^{-4} e^{2 \pi i/3}$ & 
\end{tabular}
\caption
{%
Mode amplitudes  for our two sets of initial conditions.
}
\label{tab:IC}
\end{table*}

We note that our decomposition of the angular metric (\ref{eq:angulardecomp}) and use of only traceless tensor harmonics is different from 
our previous study of the AdS$_4$ superradiant instability \cite{Chesler:2018txn}.  In particular, in Ref.~\cite{Chesler:2018txn} we expanded 
$\hat g_{ab}$ in a basis of 
\textit{three} tensor harmonics, with the additional one being proportional to $\eta_{ab}$, and imposed the constraint (\ref{eq:deteq}) numerically during each time step.  
However, imposing the constraint (\ref{eq:deteq}) involves nonlinear operations which can themselves numerically excite high angular momentum modes.  
While this can be ameliorated by employing a stronger filter than that used in this paper, the cost of doing so is reducing the effective angular resolution for 
a given value of $\ell_{\rm max}$.  Hence, while our value of $\ell_{\rm max}$ is essentially the same as that used in Ref.~\cite{Chesler:2018txn}, our effective angular resolution is superior.

\begin{figure*}[ht]
	\begin{center}
		\includegraphics[trim= 0 0 0 0,clip,scale=0.5]{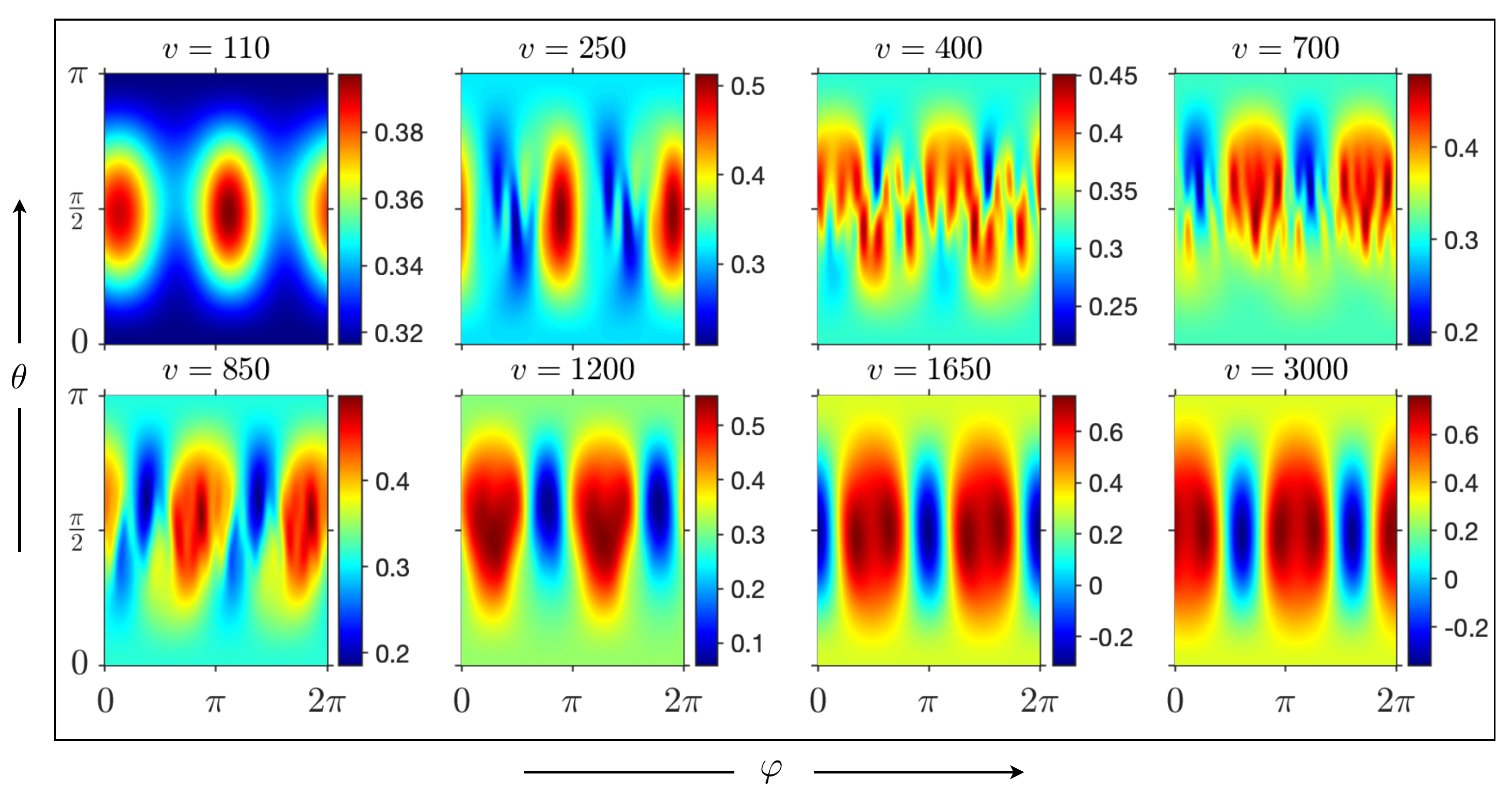}
		\caption{The boundary energy density $T^{00}$ at eight different times for I.C. A.  
		By $v = 110$ a small amplitude $m = 2$ mode is excited.
                 At subsequent times more structure develops via the excitation of higher $m$ modes.  These modes
                 rotate at different angular velocities.
                 However, by time $v = 1650$ much of this structure has relaxed.
                 At late times the energy density is approximately that of a black resonator, rigidly rotating in $\varphi$ at constant angular velocity.
                 Indeed, the energy density at $v = 1650$ and $v = 3000$ mostly differs by a rotation in $\varphi$.
                 Additionally, at late times there are also small amplitude $m = 2,4$ and 6 modes rotating at different angular velocities.
                 These  modes are responsible for the small differences seen in the structure of the energy density peaks at $v = 1650$ and $v = 3000$.
                 Note the appearance of negative energy density $v \geq 1650$.                
		}
		\label{fig:Energy}
	\end{center}
\end{figure*}

\subsection{Initial data}

The Kerr-AdS solution is parameterized by mass and spin parameters $M$ and $a$, 
and the AdS radius $L$, which we have set to unity.  
For initial data we choose 
\begin{subequations}
	\begin{eqnarray}
	h_{ab} &=& h_{ab}^{\rm Kerr},
	\\
	T_{0\mu} &=& T_{0\mu}^{\rm Kerr}+ \Delta T_{0\mu},
	\end{eqnarray}
\end{subequations}
where the superscript ``Kerr" refers to the metric and boundary stress of the Kerr-AdS solution. 
If $\Delta T^{0 \mu} = 0$, then the resulting geometry is exactly that of the Kerr-AdS solution.  Hence, with these initial conditions
the superradiant instability is seeded by non-vanishing $\Delta T^{0 \mu}$.
We choose
\begin{subequations}
\begin{eqnarray}
\Delta T_{00} &=& -\frac{4}{3}{\rm Re} \sum_{\ell = 2}^4 \alpha_{\ell} \, y^{\ell \ell},
\\
\Delta T_{0a} &=& -{\rm Re} \sum_{\ell = 2}^4 \alpha_{\ell}  \, (\mathcal V^{1 \ell \ell}_a + \mathcal V^{2\ell \ell}_a).
\end{eqnarray}
\end{subequations}
Therefore, our initial conditions are determined by the mass parameter $M$, spin parameter $a$, and the mode amplitudes $\alpha_\ell$.
Note that our choice of $\Delta T^{0\mu}$ yields vanishing contribution to the total mass and angular momentum of the system.
Hence the mass and angular momentum are identical to that of the Kerr-AdS solution with mass parameter $M$ and spin parameter $a$.

We employ mass and spin parameters
\begin{align}
&M = 0.2375,&
&a = 0.2177,&
\end{align}
and two sets of mode amplitudes $\alpha_\ell$ given in Table~\ref{tab:IC}, which 
we will refer to as initial condition (I.C.) A and B.
Note our chosen masses and spins differ from those used in our previous work \cite{Chesler:2018txn} by $\sim 5\%$.

\begin{figure*}[ht]
   \begin{center}
     \includegraphics[trim= 0 0 0 0,clip,scale=0.5]{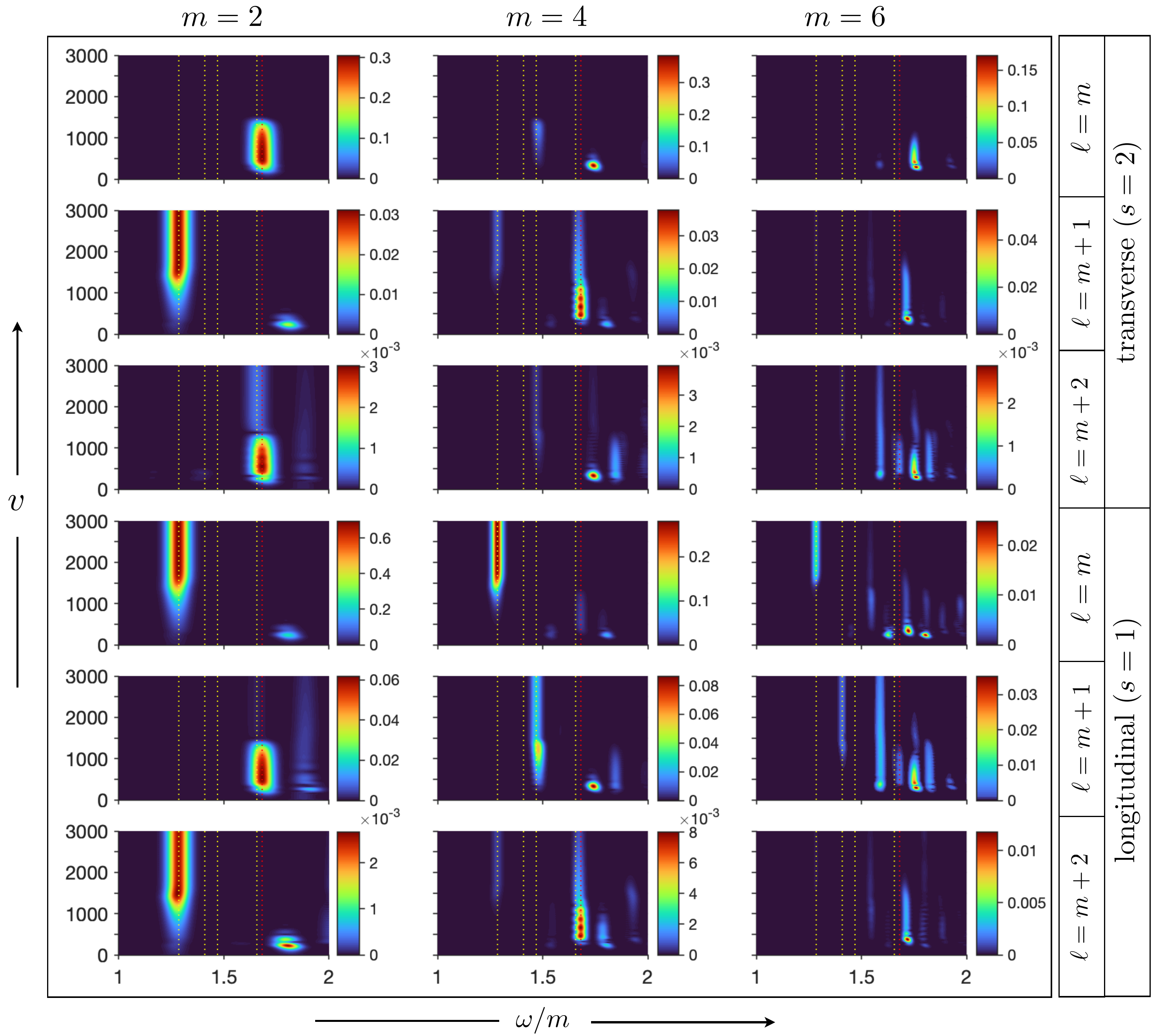}
     \caption{Spectrograms showing $|\mathcal F^{s \ell m}|$ as a function of $v$ and $\omega/m$ for I.C. A.  All plots are on the same scale.
     At sufficiently late times the mode amplitudes $|\mathcal F^{s\ell m}|$ decay everywhere except 
     in the vicinity of a few fundamental frequencies, which are denoted by the yellow dashed lines.
     The dashed red line shows a fundamental frequency of the intermediate state.  This mode is excited 
     by the initial instability, which occurs in the $(s \ell m) = (222)$ channel.
     }
     \label{fig:spectrogram}
   \end{center}
 \end{figure*}

\section{Results}
\label{sec:results}

We begin by showcasing results for I.C. A.  
In the top panel of Fig.~\ref{fig:Energy} we plot snapshots of boundary energy density $T^{00}$ at eight times between $v = 110$ and $v = 3000$.
Note that the color scaling is different at each time.  
By $v = 110$ a small amplitude $m = 2$ excitation is visible.  
At subsequent times more structure develops via the excitation of higher $m$ modes, with different modes rotating in $\varphi$ at different angular velocities.  
Nevertheless, by  $v = 1650$ most of this structure has relaxed and the energy density approximately rotates rigidly at constant angular velocity.
As we elaborate on further below, there are also tiny excitations with $m = 2,4,6$ rotating at different angular velocities.  
These tiny excitations are responsible for small differences seen in the energy density peaks at times 
$v = 1650$ and $v = 3000$.

To quantify the growth of different modes, including their frequency content, we define the mode amplitudes 
\begin{equation}
{\mathcal{F}}^{s\ell m}(v,\omega) \equiv \int d^2 \bm x dv'  e^{i \omega v'} \mathcal V^{*s \ell m}_a(\bm x) W(v{-}v') T^{0a}(v',\bm x) ,
\end{equation}
where $W(v)$ is a Gaussian window function with width 15, and the vector spherical harmonics $\mathcal V^{s \ell m}_a$ are given by Eq.~(\ref{eq:vectorharmonics}).
$\mathcal F^{1\ell m}$ and $\mathcal F^{2\ell m}$ are essentially the Fourier space amplitudes of the longitudinal and
transverse components of the momentum density, 
respectively.  More precisely, $\mathcal F^{s\ell m}$ is the 
short-time Fourier transform of the vector spherical harmonic transform of the momentum density.

\begin{figure*}[ht]
   \begin{center}
     \includegraphics[trim= 0 0 0 0,clip,scale=0.7]{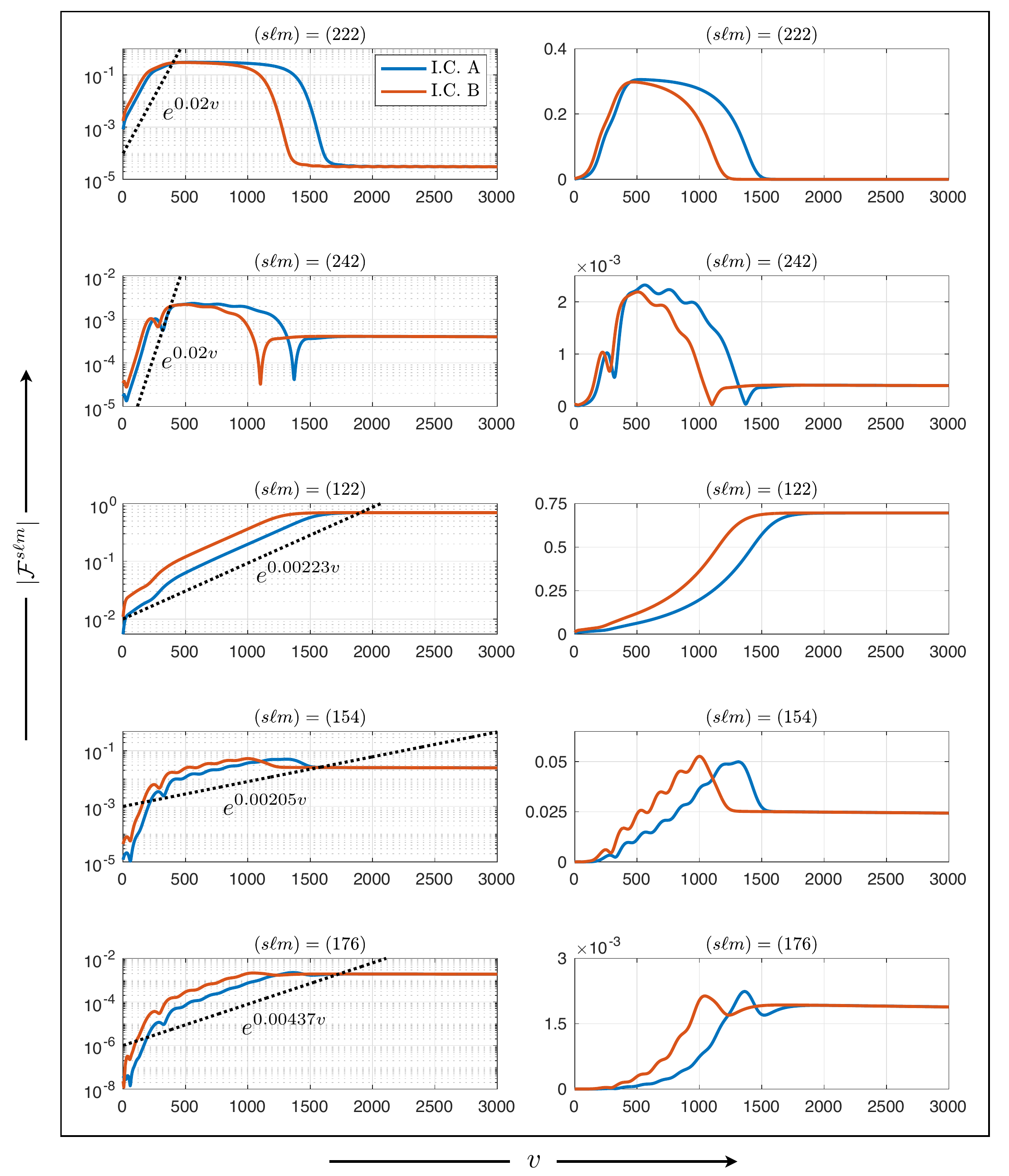}
     \caption{
        The mode amplitudes $|\mathcal F^{s  \ell m}(v,\w_{s \ell m})|$ evaluated
        along the dashed yellow and red lines shown in Fig.~\ref{fig:spectrogram}.
        The blue curves correspond to I.C. A while the red curves correspond to I.C. B.
        The left and right columns show identical data, with the left column 
        employing a logarithmic scale and the right column using a linear scale.
        Evolution arising from both sets of initial conditions exhibits both initial and secondary instabilities.
        The initial instability has $|\mathcal F^{222}| \sim e^{0.02 v}$.  After $|\mathcal F^{222}|$ 
        plateaus, the secondary instability kicks in with 
        $| \mathcal F^{122}| \sim e^{0.0022 v}$,
$|\mathcal F^{154}| \sim e^{0.0021 v}$ and 
$|\mathcal F^{176}| \sim e^{0.0041 v}.$
       After the secondary instability terminates, 
       the mode amplitudes approach constants, which are the same for both sets of initial conditions.
}
     \label{fig:modeamps45}
   \end{center}
 \end{figure*}

In all of our simulations we see no significant growth in $\mathcal F^{s \ell m}$ with $m = 0$ or odd $m$.
We therefore focus on even $m$.
Plotted in Fig.~\ref{fig:spectrogram} are spectrograms showing $|{\mathcal{F}}^{s\ell m}|$ 
as a function of $v$ and $\omega/m$ for $m = 2,4,6$
and $\ell = m,m+1,m+2$.  All plots are on the same scale.  Note $\omega/m$ is the angular velocity of a 
mode with frequency $\omega$ and azimuthal quantum number $m$.
At sufficiently late times the mode amplitudes $|\mathcal F^{s\ell m}|$ decay everywhere except 
in the vicinity of a few fundamental frequencies.  The four yellow dashed lines in Fig.~\ref{fig:spectrogram}
lie at frequencies $\w_{s \ell m}$, where $|\mathcal F^{s \ell m}(v,\w_{s \ell m})|$ is approximately constant at late times.
These angular velocities are
\begin{align}
\label{eq:fundamentals}
&\textstyle \w_{122} = 1.29,& 
&\textstyle \w_{176} = 1.41.&
&\textstyle \w_{154} = 1.47,&
&\textstyle \w_{242} = 1.65. &
\end{align}
Additionally, the red dashed line shows 
\begin{equation}
\label{eq:fundamentals0}
\textstyle \w_{222} = 1.68.
\end{equation}
The mode amplitudes $|\mathcal F^{s \ell m}(v,\w_{s \ell m})|$ evaluated along the dashed curves are shown below in Fig.~\ref{fig:modeamps45}.
Note that the left and right columns in Fig.~\ref{fig:modeamps45} show identical data, with the left column employing a logarithmic scale while the right column uses 
a linear scale.

 \begin{figure*}
	\begin{center}
		\includegraphics[trim= 0 0 0 0,clip,scale=0.4]{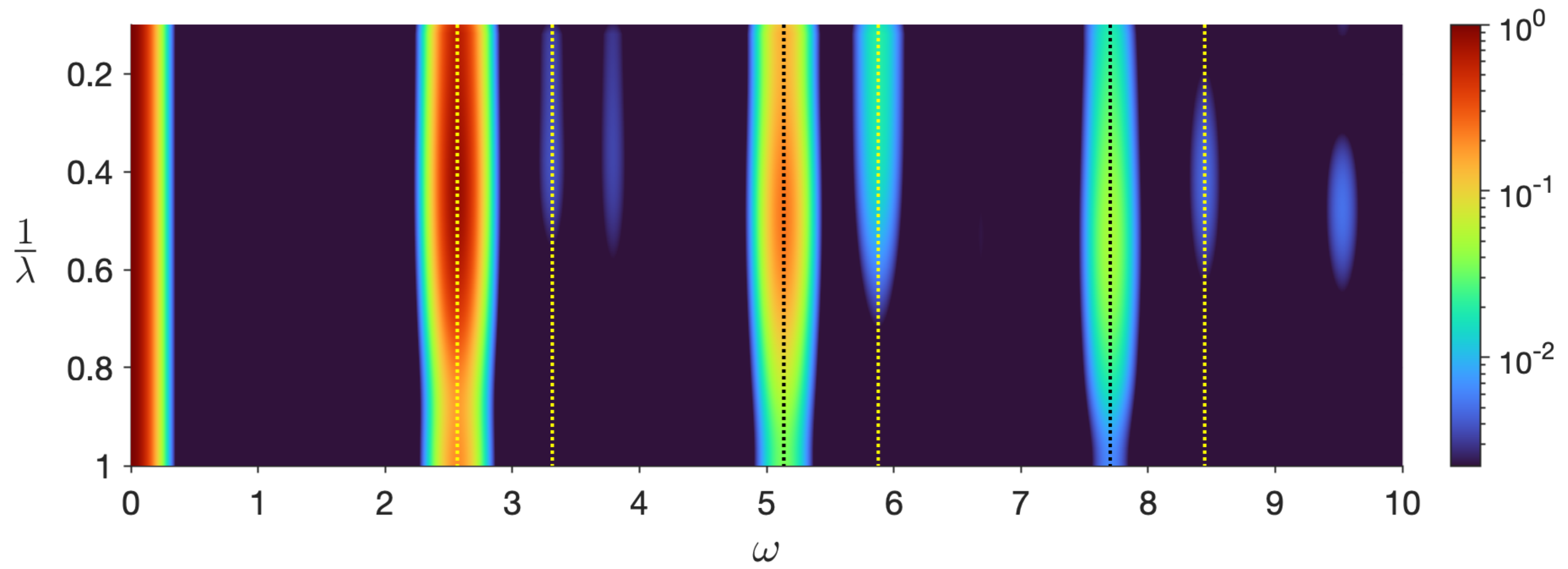}
		\caption{$\mathscr R$ as a function of inverse radius $1/\lambda$ and frequency $\omega$ for I.C. A.
		The four yellow lines superimposed on the figure
                 correspond to  the four angular velocities $\omega/m = \w_{s \ell m}$ shown in Fig.~\ref{fig:spectrogram}.
		The black lines show harmonics of the black resonator fundamental frequency, $\omega = 2 \w_{122} = 2.59$.   
		Note that the color scaling is on a logarithmic scale.  
		$\mathscr R$ is peaked at several discrete frequencies,
                 with the largest peaks at $\omega = 0$ and the black resonator's fundamental frequency.
                 Modes corresponding to the black resonator's hair are largest at large radii.
		}
		\label{fig:curvatureplot}
	\end{center}
\end{figure*}

From Figs.~\ref{fig:spectrogram} and \ref{fig:modeamps45} it is evident there are two 
distinct epochs of growth in $|\mathcal F^{s \ell m}|$, which we shall refer to as initial and secondary instabilities \cite{Chesler:2018txn}.
The initial instability is due to superradiant instabilities in 
the Kerr-AdS geometry.  Our numerics are consistent with 
an initial instability in the $(s\ell m) = (222)$ channel with complex frequency
\begin{equation}
\label{eq:KerrAdSFreq}
\omega_{222} = 3.41 + 0.020 i.
\end{equation}
The growth of $|\mathcal F^{222}|$ begins to slow at 
$v \sim 300$, while also \textit{slightly} shifting in frequency, and reaches its peak amplitude around $v \sim 500$.
During the subsequent interval, $500 \lesssim v \lesssim 1200$, there is a plateau-like structure in $|\mathcal F^{222}(v,\w_{222})|$.  
It is also during this time interval that the secondary instability kicks in, with  
\begin{align}
\label{eq:secondarygrowth}
&| \mathcal F^{122}| \sim e^{0.0022 v},&
&|\mathcal F^{154}| \sim e^{0.0021 v},&
&|\mathcal F^{176}| \sim e^{0.0041 v}.&
\end{align}
Note that the exponential growth of $|\mathcal F^{154}| $ and $|\mathcal F^{176}|$ is 
not as crisp as that of $|\mathcal F^{122}|$.

The fact that $|\mathcal F^{222}|$ is the largest amplitude mode at the early stages of the secondary instability 
suggests that the intermediate state is an excited black resonator.  However, it is noteworthy that near $v \sim 500$ there
are additional modes excited with amplitudes comparable to $|\mathcal F^{222}|$, but with 
fundamental frequencies different from $\w_{222}$.  See Fig.~\ref{fig:spectrogram}.
We comment on this further below in the Discussion section.

\begin{figure*}
	\begin{center}
		\includegraphics[trim= 0 0 0 0,clip,scale=0.4]{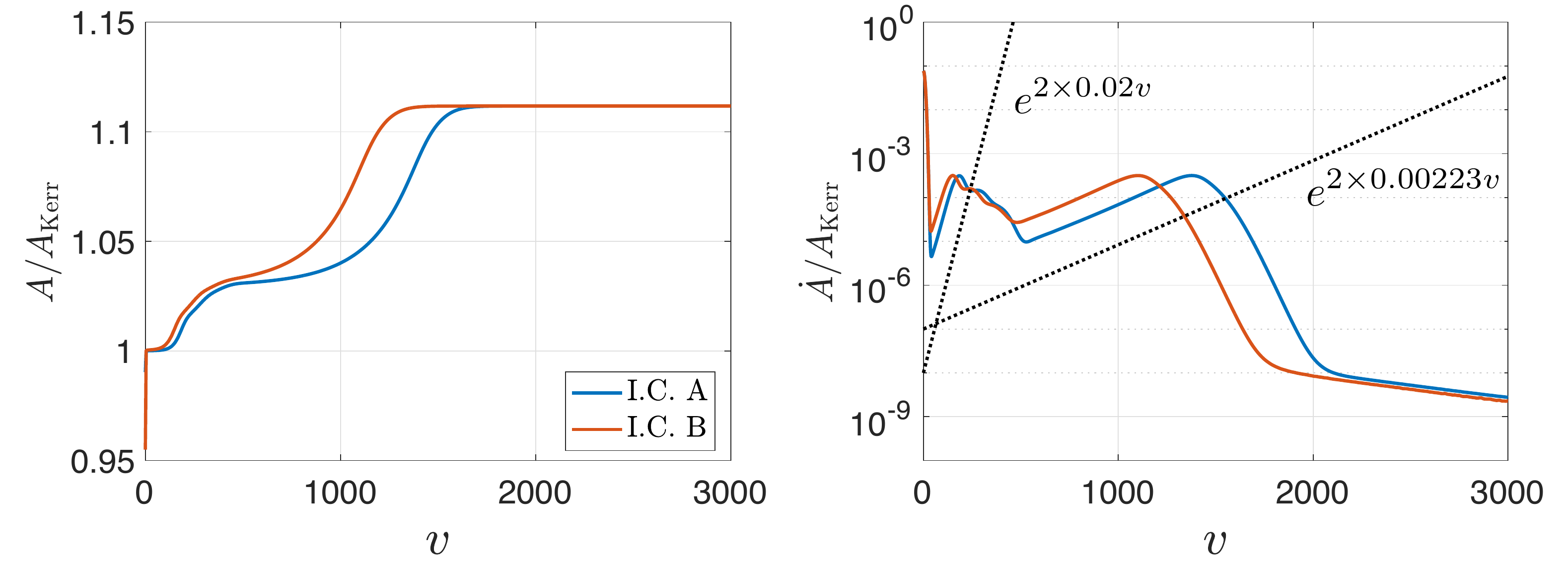}
		\caption{The apparent horizon area $A$ (left) and the rate of area growth $\dot A \equiv \partial_v A$ (right) for I.C. A (blue) and I.C. B (red).
		The areas are normalized 
		by the Kerr-AdS horizon area.  The dashed black lines in the right panel are the squares of the two exponential curves 
		seen in the $(s \ell m) = (222)$ and $(s \ell m) = (122)$
                 panels of Fig.~\ref{fig:modeamps45}.
		For both sets of initial conditions, two distinct epochs of horizon area growth are clearly visible.
		At late times $\dot A$ decreases exponentially.
		At $v = 3000$ the two apparent horizon areas agree at order 1 part in $10^6.$
		}
		\label{fig:Area}
	\end{center}
\end{figure*}
  
As seen in both Figs.~\ref{fig:spectrogram} and \ref{fig:modeamps45}, 
after $v \sim 1200$, $|\mathcal F^{222}|$ begins to precipitously decay
while $|\mathcal F^{242}|$, $|\mathcal F^{122}|$, $|\mathcal F^{154}|$ and $|\mathcal F^{176}|$
approach constants shortly thereafter.
Actually, all three of these modes are decaying extremely slowly 
at late times, with $\partial_v \log | \mathcal F^{s \ell m} | \sim -1.9 \times 10^{-5}$, although 
this is nearly impossible to see in Fig.~\ref{fig:modeamps45}.  We are unsure if this tiny decay rate is real or 
a numerical artifact.
We comment on this below in the Discussion section.
At $v = 3000$ we have,
\begin{subequations}
\begin{align}
&|\mathcal F^{122}| \approx 0.70,&
&|\mathcal F^{154}| \approx 0.024.&
\\
&|\mathcal F^{172}| \approx 1.9 \times 10^{-3},&
&|\mathcal F^{242}| \approx 4 \times 10^{-4}.&
\end{align}
\end{subequations}
It is noteworthy that $|\mathcal F^{122}|$ is nearly 30 times larger than the next 
largest mode, $|\mathcal F^{154}|$.  
If $\w_{122}$ was
the only angular velocity excited, then the geometry would be a black resonator.
Evidently, the final state of our numerical evolution is a black resonator
with a small amount of gravitational hair.  

A natural question is where in the bulk is the black resonator's hair localized? 
Is it localized near the boundary or is it spread throughout the bulk, including near the horizon? 
To answer this question we consider the Kretschmann scalar
$K \equiv R^{\mu \nu \alpha \beta}R_{\mu \nu \alpha \beta}$.  Near the boundary $K \to 24$, indicating that $K$ itself 
is not a good probe of near-boundary excitations.  We wish to construct a quantity from $K$ which has a nontrivial limit near the boundary
and which can discriminate contributions from different 
frequencies, including those with small mode amplitudes.
To this end we take the short-time Fourier transform of the difference $\Delta K \equiv K - 24$,%
\footnote{Due to memory constraints relevant for the calculation of $K$, here we use a Gaussian window function $W(v)$ with width 10.}
\begin{equation}
\Delta\widetilde K(v,\omega,\lambda,\bm x)   \equiv  \int dv' e^{i \omega v'} W(v - v') \Delta K(v,\lambda,\bm x),
\end{equation}
and then average $|\Delta \widetilde K|$ over the angular directions,
\begin{equation}
\langle | \Delta \widetilde K(v,\omega,\lambda) | \rangle \equiv \int d^2 \bm x \, |\Delta \widetilde K(v,\omega,\lambda,\bm x)|.
\end{equation}
$\langle | \Delta \widetilde K(v,\omega,\lambda) | \rangle$  vanishes near the boundary and is insensitive to the precise value of $v$ (at sufficiently late times).  
The dimensionless ratio 
\begin{equation}
\label{eq:Rdef}
\mathscr R \equiv \frac{ \langle | \Delta \widetilde K| \rangle}{\langle | \Delta \widetilde K| \rangle|_{\omega = 0}},
\end{equation} 
has a nontrivial boundary limit and 
measures the amplitude of a bulk mode relative to the zero frequency limit.

In Fig.~\ref{fig:curvatureplot} we plot $\mathscr R$ at time $v = 2750$
as a function of the radial coordinate $\lambda$ and frequency $\omega$.
The four yellow lines superimposed on the figure
correspond to  the four angular velocities $\omega/m = \w_{s \ell m}$ shown in Fig.~\ref{fig:spectrogram} and  listed in Eq.~\ref{eq:fundamentals}.
The black lines show harmonics of the black resonator fundamental frequency, $\omega = 2 \w_{122} = 2.59$.   Note that the color scaling is on a logarithmic scale.
As is evident from the figure, $\mathscr R$ is peaked at several discrete frequencies,
with the largest peaks at $\omega = 0$ and the black resonator's fundamental frequency.
The peaks corresponding to the black resonator's hair
are largest at large radii and smallest at the horizon ($\lambda = 1$).  In particular, at the horizon 
$\mathscr R < 0.4 \times 10^{-4}$ everywhere except near $\omega = 0$ and the black resonator's 
fundamental frequency and harmonics.  In contrast, 
at $\lambda = 5$, the peak at $\omega = 4 \w_{154}$ has amplitude  
$0.02$.  This signals that the black resonator's hair is concentrated away from the horizon.
Similar behavior was seen in Ref.~\cite{Bosch:2016vcp}, where the final state of the (spherically symmetric)
charged AdS superradiant instability was studied.

Plotted in Fig.~\ref{fig:Area} 
is the apparent horizon area $A$ (left panel) and its time derivative $\dot A \equiv \partial_v A$ (right panel), both normalized by the area of the associated Kerr-AdS black hole.
To increase the fidelity of the plots of $\dot A$, as well as smooth out short time structure in $\dot A$ due to 
individual superradiant amplification cycles,
we compute $\dot A$ by convolving $A$ with the time derivative of a normalized Gaussian with width 8.
The two dashed lines in the right panel are simply the squares of the exponential curves seen in the $(s \ell m) = (222)$ and $(s \ell m) = (122)$
panels of Fig.~\ref{fig:modeamps45}.
Evidently, during both the initial and secondary instabilities, the horizon area grows roughly as the square of the these modes.  
Note that $\dot A$ decreases exponentially at late times.  At $v = 3000$ the horizon area is  
11\% larger than that of the Kerr-AdS black hole with the same mass and spin.  

Finally, we turn to evolution generated by our second initial condition.  I.C. B has the same energy and angular momentum as I.C. A, but different initial 
seed perturbations.  It turns out that many of the features of the resulting solutions are identical, both qualitatively and quantitatively.
The resulting spectrogram for I.C. B looks strikingly similar to that shown in Fig.~\ref{fig:spectrogram}, with angular velocities $\w_{s \ell m}$  identical to 
those found with I.C. A.  Also included in Fig.~\ref{fig:modeamps45}  are the amplitudes $|\mathcal F^{s \ell m}(v,\w_{s \ell m})|$ obtained with I.C. B.
During the initial and secondary instabilities the exponential growth rates are the same for both sets of initial conditions.
Likewise, as shown in Fig.~\ref{fig:Area}, $A$ and $\dot A$ also have similar structure for both sets of initial conditions.
Remarkably, the final values of $|\mathcal F^{122}|$,  $|\mathcal F^{242}|$, $|\mathcal F^{154}|$ and $|\mathcal F^{176}|$ 
are also nearly identical for both set of initial conditions.  Moreover, the final apparent horizon areas 
agree at order 1 part in $10^{6}$.

Despite these similarities, the final state obtained with I.C. B is in fact distinct from that obtained with I.C. A.
One difference lies in relative phase shifts between different modes.  Define the phases
\begin{align}
&\delta_{s \ell m}= {\rm arg} \left( \mathcal F^{s \ell m} \right ),& & {\rm with}&
&\delta_{s \ell m} \in [-\pi,\pi].&
\end{align}
In order the exclude the possibility that the phases $\delta_{s \ell m}$ of our two solutions 
are merely related by a shift in $v$ and/or $\varphi$, we define
the weighted phase differences
\begin{eqnarray}
\Delta_1 &=& 8  (\w_{154}-\w_{242}) \delta_{122}+4  (\w_{242}-\w_{122}) \delta_{154}
 \\ \nonumber
 &+&8(\w_{122}-\w_{154}) \delta_{242},
\\
\Delta_2 &=& 24 (\w_{154}-\w_{176}) \delta_{122} +12 (\w_{176}-\w_{122})  \delta_{154} \ \ \ \ \ \ \ \ \ 
\\ \nonumber
&+&8 (\w_{122}-\w_{154}) \delta_{176}. 
\end{eqnarray}
It is easy to check that $\Delta_1$ and $\Delta_2$ are invariant under 
shifts in $v$ and $\varphi$.  If the late-time solutions generated by 
different initial conditions are identical, then 
they must have identical $\Delta_1$ and $\Delta_2$ at late times.

\begin{figure*}
	\begin{center}
		\includegraphics[trim= 0 0 0 0,clip,scale=0.45]{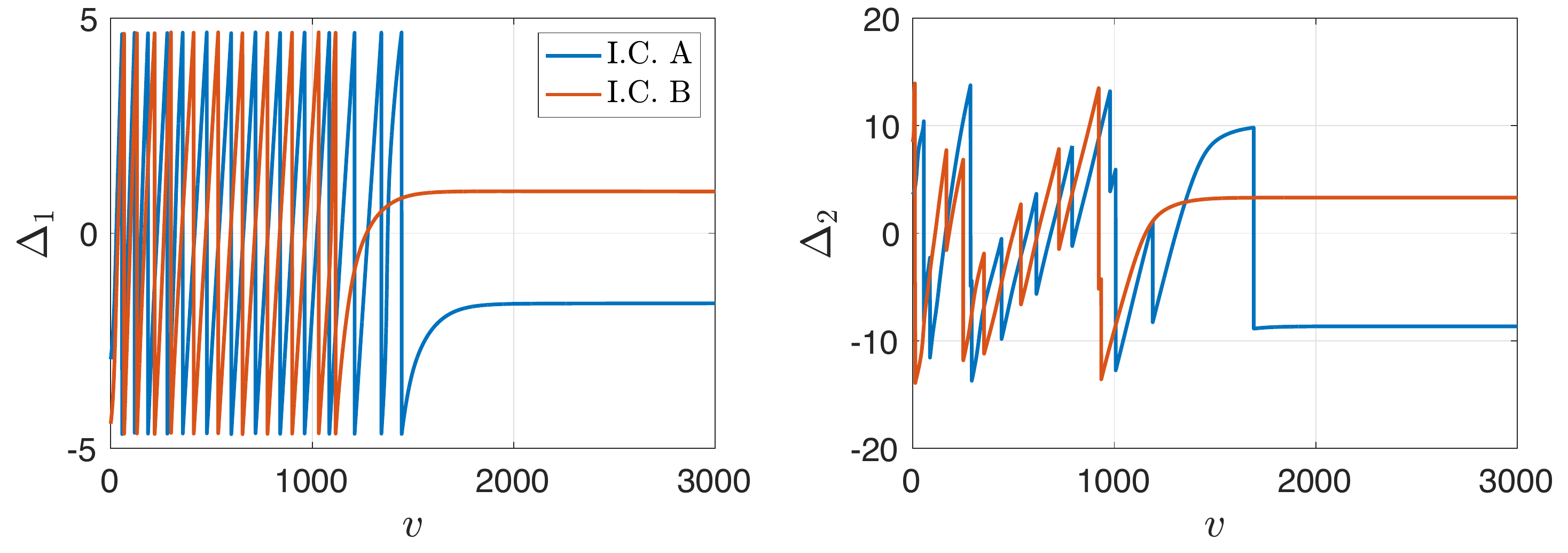}
		\caption{Weighted phase differences $\Delta_1$ and $\Delta_2$ for both sets of initial conditions.
		The saw-tooth like structure seen in both plots is due to the fact that the phases $\delta_{s \ell m}$
		are not continuous functions of time.  For both 
                sets of initial conditions $\Delta_1$ and $\Delta_2$ approach constants at late times.
                However, these constants are different for I.C. A and I.C. B.  This demonstrates 
                that the gravitational hair obtained with I.C. A is distinct from that generated with I.C. B.
		}
		\label{fig:phaseplot}
	\end{center}
\end{figure*}

In Fig.~\ref{fig:phaseplot} we plot $\Delta_1$ and $\Delta_2$.  The saw-tooth like structure seen in both plots 
is due to the fact that the phases $\delta_{s \ell m}$
are not continuous functions of time.  For both 
sets of initial conditions $\Delta_1$ and $\Delta_2$ approach constants at late times.
However, these constants are different for each initial condition.  This demonstrates 
that the gravitational hair obtained with I.C. A is distinct from that generated with I.C. B.
Simply put, the final state of our evolution is sensitive to seed perturbation in the Kerr-AdS geometry.

\section{Discussion}
\label{sec:discussion}

It is remarkable that after starting in the Kerr-AdS geometry, and then having many large $m$ modes excited --- 
as is evident from the plots of the energy density seen in Fig.~\ref{fig:Energy} ---
that the geometry finds its way to a hairy black resonator state.  
Is a hairy black resonator, or more generally a multi-oscillating black hole, the final state of the superradiant instability?
While our numerics are certainly consistent with this ---  mode amplitudes and the horizon area appear to plateau --- we cannot exclude other possibilities.
For example, it is always possible there are additional 
instabilities which are too small to be observed with our current numerical simulations. Likewise, its always possible there are instabilities in
other channels not probed by our limited set of initial conditions.  

Even if the final state is a hairy black resonator, it is possible the hair could undergo nontrivial 
slow dynamics.  For example, the hair could decay or different modes could interact via nonlinear couplings, resulting 
in slow changes to the hairy mode amplitudes.
We do see some evidence of slow mode amplitude dynamics in our simulations.
The hairy modes seen in Fig.~\ref{fig:modeamps45} all are decaying very 
slowly at late times, with $\partial_v \log | \mathcal F^{s \ell m} | \sim -1.9 \times 10^{-5}$.
Additionally, we do see a tiny amount of late time growth in $| \mathcal F^{133} | \sim 3 \times 10^{-3}$, 
which roughly grows like $\partial_v \log | \mathcal F^{133} | \sim +7 \times 10^{-6}$.
This mode has angular velocity $\w_{133} = 1.27$, which only differs from $\w_{122}$ by $1.5\%$.
However, while these tiny rates could indicate interesting hairy dynamics  (or even instabilities 
of the black resonator itself), they lie at the threshold of fidelity of 
our current numerical simulations.  Indeed, we have also ran short duration simulations 
of our final state at 15\% reduced resolution and found these rates to change significantly.
To help assess whether the time dependence of these mode amplitudes is physical,
it would be useful run simulations with different discretization and filtering schemes (e.g. cubing the sphere as opposed to using spherical harmonics),
as well as long duration high resolution simulations.

We have also ran short duration simulations of a variety of perturbations of our final state.
In doing so we found that after initial transients decay, 
the hairy mode amplitudes become approximately constant, 
just as seen in Fig.~\ref{fig:modeamps45}.
However, we have found that the associated constants are 
sensitive to the perturbations and can differ by O(1) factors from those
seen in Fig.~\ref{fig:modeamps45}.  Notably however, these solutions 
have virtually identical horizon areas:
all simulations resulted in the same horizon area at order 1 part in $10^6$.  
This presumably reflects the fact that, as highlighted in Fig.~\ref{fig:curvatureplot}, the hair is localized near the boundary, 
meaning it does not affect the near-horizon geometry.  

It would clearly be of considerable value to study the mode spectrum of black resonators.
First, it would be interesting to 
see how instabilities and decay rates of the black resonator associated with our final state 
compare with our numerical simulations.  
Second, it would be interesting to compare our intermediate state to a black resonator.
While the spectrograms in Fig.~\ref{fig:spectrogram} demonstrate that the intermediate state is initially dominated by a single angular velocity, $\w_{222}$, 
there are other modes appreciably excited with different angular velocities.
This observation alone does not exclude the possibility that the intermediate state is an excited black resonator. 
Indeed, far-from-equilibrium black branes in AdS$_5$ are well-approximated by 
large amplitude linear perturbations on top of a stationary background \cite{Heller:2012km}.
To solidify the intuition that the the intermediate state is an approximate black resonator, it would be useful to 
compare the black resonator spectrum to the spectrum of decaying and growing modes seen during 
the intermediate stage of our evolution.  Finally, it would also be interesting 
to understand why instabilities associated with the intermediate 
state are strong, while none are visible in our final state. At the boundary our intermediate state is dominated by transverse modes whereas our final state is dominated by 
longitudinal modes.  It is possible that instabilities associated with these two different types of black resonators could be 
qualitatively different.

We have only studied evolution with a single set of masses and spins.
A natural question then is how generic are our results?  Is the resulting multi-oscillating black hole
always close to a black resonator?  Its noteworthy that our previous work, Ref.~\cite{Chesler:2018txn}, which studied evolution with a different mass and spin, 
resulted in $|\mathcal F^{154}| \sim |\mathcal F^{122}|$ during the intermediate stage of the evolution.
This suggests it is possible that the final state can be a multi-oscillating black hole which is significantly different from a black resonator.
Simulating such a scenario likely requires higher resolution and longer runtimes than the simulations presented in this paper.

Finally, via AdS/CFT duality our results also have interesting consequences for quantum field theory.
Firstly, the existence of the AdS superradiant instability implies that rotation invariance is spontaneously broken 
in the dual field theory.  This behavior evidently only occurs in small systems, where there is are dual superradiant instabilities.
From field theory arguments alone, we do not know why one should have expected 
this.  Second, both sets initial conditions studied in this paper yield evolution with 
negative energy density.  This was also found in Ref.~\cite{Chesler:2018txn}.
This indicates that the dual quantum field theory state
has exotic properties.  Regions of negative energy density have also been seen in the holographic 
models studied in \cite{Athanasiou:2010pv,Casalderrey-Solana:2013aba,Chesler:2015fpa,Horowitz:2014hja}.  
Finally, it is remarkable that the system does not appear to approach a stationary configuration
despite having a large entropy.  
It would be interesting to explore to what extent, if any, hairy black resonators
correspond to thermal states in the dual quantum field theory.   
This should be possible using the methods developed in Refs.~\cite{Caron-Huot:2011vtx,Chesler:2011ds}.
We leave this and many other interesting question for future work.

\begin{acknowledgments}
I thank Luis Lehner and Larry Yaffe for helpful comments and discussions.
\end{acknowledgments}

\bibliographystyle{utphys}
\bibliography{refs}%

\providecommand{\href}[2]{#2}\begingroup\raggedright\begin{thebibliography}{10}

\bibitem{zeldovich}
Y.~B. Zel'dovich, ``Generation of waves by a rotating body,'' {\em JETP Lett.}
  {\bf 14} (1971)  180.

\bibitem{Starobinsky:1973aij}
A.~A. Starobinsky, ``{Amplification of waves reflected from a rotating ''black
  hole''.},'' {\em Sov. Phys. JETP} {\bf 37} (1973) no.~1, 28--32.

\bibitem{Penrose:1969pc}
R.~Penrose, ``{Gravitational collapse: The role of general relativity},''
  \href{http://dx.doi.org/10.1023/A:1016578408204}{{\em Riv. Nuovo Cim.} {\bf
  1} (1969)  252--276}.

\bibitem{Penrose:1971uk}
R.~Penrose and R.~M. Floyd, ``{Extraction of rotational energy from a black
  hole},'' \href{http://dx.doi.org/10.1038/physci229177a0}{{\em Nature} {\bf
  229} (1971)  177--179}.

\bibitem{Press:1972zz}
W.~H. Press and S.~A. Teukolsky, ``{Floating Orbits, Superradiant Scattering
  and the Black-hole Bomb},''
\href{http://dx.doi.org/10.1038/238211a0}{{\em Nature} {\bf 238} (1972)
  211--212}.

\bibitem{Pani:2013hpa}
P.~Pani and A.~Loeb, ``{Constraining Primordial Black-Hole Bombs through
  Spectral Distortions of the Cosmic Microwave Background},''
  \href{http://dx.doi.org/10.1103/PhysRevD.88.041301}{{\em Phys. Rev.} {\bf
  D88} (2013)  041301},
\href{http://arxiv.org/abs/1307.5176}{{\tt arXiv:1307.5176 [astro-ph.CO]}}.

\bibitem{Arvanitaki:2016qwi}
A.~Arvanitaki, M.~Baryakhtar, S.~Dimopoulos, S.~Dubovsky, and R.~Lasenby,
  ``{Black Hole Mergers and the QCD Axion at Advanced LIGO},''
  \href{http://dx.doi.org/10.1103/PhysRevD.95.043001}{{\em Phys. Rev.} {\bf
  D95} (2017) no.~4, 043001},
\href{http://arxiv.org/abs/1604.03958}{{\tt arXiv:1604.03958 [hep-ph]}}.

\bibitem{East:2017ovw}
W.~E. East and F.~Pretorius, ``{Superradiant Instability and Backreaction of
  Massive Vector Fields around Kerr Black Holes},''
\href{http://arxiv.org/abs/1704.04791}{{\tt arXiv:1704.04791 [gr-qc]}}.

\bibitem{East:2018glu}
W.~E. East, ``{Massive Boson Superradiant Instability of Black Holes: Nonlinear
  Growth, Saturation, and Gravitational Radiation},''
  \href{http://dx.doi.org/10.1103/PhysRevLett.121.131104}{{\em Phys. Rev.
  Lett.} {\bf 121} (2018) no.~13, 131104},
  \href{http://arxiv.org/abs/1807.00043}{{\tt arXiv:1807.00043 [gr-qc]}}.

\bibitem{Siemonsen:2019ebd}
N.~Siemonsen and W.~E. East, ``{Gravitational wave signatures of ultralight
  vector bosons from black hole superradiance},''
  \href{http://dx.doi.org/10.1103/PhysRevD.101.024019}{{\em Phys. Rev. D} {\bf
  101} (2020) no.~2, 024019}, \href{http://arxiv.org/abs/1910.09476}{{\tt
  arXiv:1910.09476 [gr-qc]}}.

\bibitem{Brito:2015oca}
R.~Brito, V.~Cardoso, and P.~Pani, ``{Superradiance},''
  \href{http://dx.doi.org/10.1007/978-3-319-19000-6}{{\em Lect. Notes Phys.}
  {\bf 906} (2015)  pp.1--237},
\href{http://arxiv.org/abs/1501.06570}{{\tt arXiv:1501.06570 [gr-qc]}}.

\bibitem{Hartnoll:2011fn}
S.~A. Hartnoll, ``{Horizons, holography and condensed matter},'' in {\em Black
  holes in higher dimensions}, G.~T. Horowitz, ed., pp.~387--419.
\newblock 2012.
\newblock \href{http://arxiv.org/abs/1106.4324}{{\tt arXiv:1106.4324
  [hep-th]}}.
\newblock
\url{http://inspirehep.net/record/914550/files/arXiv:1106.4324.pdf}.
\newblock

\bibitem{Chesler:2014gya}
P.~M. Chesler, A.~M. Garcia-Garcia, and H.~Liu, ``{Defect Formation beyond
  Kibble-Zurek Mechanism and Holography},''
  \href{http://dx.doi.org/10.1103/PhysRevX.5.021015}{{\em Phys. Rev. X} {\bf 5}
  (2015) no.~2, 021015}, \href{http://arxiv.org/abs/1407.1862}{{\tt
  arXiv:1407.1862 [hep-th]}}.

\bibitem{Maldacena:1997re}
J.~M. Maldacena, ``{The Large N limit of superconformal field theories and
  supergravity},'' \href{http://dx.doi.org/10.1023/A:1026654312961,
  10.4310/ATMP.1998.v2.n2.a1}{{\em Int. J. Theor. Phys.} {\bf 38} (1999)
  1113--1133}, \href{http://arxiv.org/abs/hep-th/9711200}{{\tt
  arXiv:hep-th/9711200 [hep-th]}}.
[Adv. Theor. Math. Phys.2,231(1998)].

\bibitem{Hawking:1999dp}
S.~W. Hawking and H.~S. Reall, ``{Charged and rotating AdS black holes and
  their CFT duals},'' \href{http://dx.doi.org/10.1103/PhysRevD.61.024014}{{\em
  Phys. Rev. D} {\bf 61} (2000)  024014},
  \href{http://arxiv.org/abs/hep-th/9908109}{{\tt arXiv:hep-th/9908109}}.

\bibitem{Cardoso:2004hs}
V.~Cardoso and O.~J.~C. Dias, ``{Small Kerr-anti-de Sitter black holes are
  unstable},'' \href{http://dx.doi.org/10.1103/PhysRevD.70.084011}{{\em Phys.
  Rev.} {\bf D70} (2004)  084011},
\href{http://arxiv.org/abs/hep-th/0405006}{{\tt arXiv:hep-th/0405006
  [hep-th]}}.

\bibitem{Cardoso:2013pza}
V.~Cardoso, O.~J.~C. Dias, G.~S. Hartnett, L.~Lehner, and J.~E. Santos,
  ``{Holographic thermalization, quasinormal modes and superradiance in
  Kerr-AdS},'' \href{http://dx.doi.org/10.1007/JHEP04(2014)183}{{\em JHEP} {\bf
  04} (2014)  183},
\href{http://arxiv.org/abs/1312.5323}{{\tt arXiv:1312.5323 [hep-th]}}.

\bibitem{Green:2015kur}
S.~R. Green, S.~Hollands, A.~Ishibashi, and R.~M. Wald, ``{Superradiant
  instabilities of asymptotically anti-de Sitter black holes},''
  \href{http://dx.doi.org/10.1088/0264-9381/33/12/125022}{{\em Class. Quant.
  Grav.} {\bf 33} (2016) no.~12, 125022},
\href{http://arxiv.org/abs/1512.02644}{{\tt arXiv:1512.02644 [gr-qc]}}.

\bibitem{Niehoff:2015oga}
B.~E. Niehoff, J.~E. Santos, and B.~Way, ``{Towards a violation of cosmic
  censorship},'' \href{http://dx.doi.org/10.1088/0264-9381/33/18/185012}{{\em
  Class. Quant. Grav.} {\bf 33} (2016) no.~18, 185012},
\href{http://arxiv.org/abs/1510.00709}{{\tt arXiv:1510.00709 [hep-th]}}.

\bibitem{Dias:2015rxy}
{\'O}.~J.~C. Dias, J.~E. Santos, and B.~Way, ``{Black holes with a single
  Killing vector field: black resonators},''
  \href{http://dx.doi.org/10.1007/JHEP12(2015)171}{{\em JHEP} {\bf 12} (2015)
  171},
\href{http://arxiv.org/abs/1505.04793}{{\tt arXiv:1505.04793 [hep-th]}}.

\bibitem{Chesler:2018txn}
P.~M. Chesler and D.~A. Lowe, ``{Nonlinear Evolution of the AdS$_4$
  Superradiant Instability},''
  \href{http://dx.doi.org/10.1103/PhysRevLett.122.181101}{{\em Phys. Rev.
  Lett.} {\bf 122} (2019) no.~18, 181101},
  \href{http://arxiv.org/abs/1801.09711}{{\tt arXiv:1801.09711 [gr-qc]}}.

\bibitem{Choptuik:2019zji}
M.~Choptuik, R.~Masachs, and B.~Way, ``{Multioscillating Boson Stars},''
  \href{http://dx.doi.org/10.1103/PhysRevLett.123.131101}{{\em Phys. Rev.
  Lett.} {\bf 123} (2019) no.~13, 131101},
  \href{http://arxiv.org/abs/1904.02168}{{\tt arXiv:1904.02168 [gr-qc]}}.

\bibitem{Ishii:2021xmn}
T.~Ishii, K.~Murata, J.~E. Santos, and B.~Way, ``{Multioscillating black
  holes},'' \href{http://dx.doi.org/10.1007/JHEP05(2021)011}{{\em JHEP} {\bf
  05} (2021)  011}, \href{http://arxiv.org/abs/2101.06325}{{\tt
  arXiv:2101.06325 [hep-th]}}.

\bibitem{Sanchis-Gual:2015lje}
N.~Sanchis-Gual, J.~C. Degollado, P.~J. Montero, J.~A. Font, and C.~Herdeiro,
  ``{Explosion and Final State of an Unstable Reissner-Nordström Black
  Hole},'' \href{http://dx.doi.org/10.1103/PhysRevLett.116.141101}{{\em Phys.
  Rev. Lett.} {\bf 116} (2016) no.~14, 141101},
\href{http://arxiv.org/abs/1512.05358}{{\tt arXiv:1512.05358 [gr-qc]}}.

\bibitem{Bosch:2016vcp}
P.~Bosch, S.~R. Green, and L.~Lehner, ``{Nonlinear Evolution and Final Fate of
  Charged Anti--de Sitter Black Hole Superradiant Instability},''
  \href{http://dx.doi.org/10.1103/PhysRevLett.116.141102}{{\em Phys. Rev.
  Lett.} {\bf 116} (2016) no.~14, 141102},
\href{http://arxiv.org/abs/1601.01384}{{\tt arXiv:1601.01384 [gr-qc]}}.

\bibitem{Bosch:2019anc}
P.~Bosch, S.~R. Green, L.~Lehner, and H.~Roussille, ``{Excited hairy black
  holes: Dynamical construction and level transitions},''
  \href{http://dx.doi.org/10.1103/PhysRevD.102.044014}{{\em Phys. Rev. D} {\bf
  102} (2020) no.~4, 044014}, \href{http://arxiv.org/abs/1912.05598}{{\tt
  arXiv:1912.05598 [gr-qc]}}.

\bibitem{Chesler:2008hg}
P.~M. Chesler and L.~G. Yaffe, ``{Horizon formation and far-from-equilibrium
  isotropization in supersymmetric Yang-Mills plasma},''
  \href{http://dx.doi.org/10.1103/PhysRevLett.102.211601}{{\em Phys. Rev.
  Lett.} {\bf 102} (2009)  211601}, \href{http://arxiv.org/abs/0812.2053}{{\tt
  arXiv:0812.2053 [hep-th]}}.

\bibitem{Chesler:2009cy}
P.~M. Chesler and L.~G. Yaffe, ``{Boost invariant flow, black hole formation,
  and far-from-equilibrium dynamics in N = 4 supersymmetric Yang-Mills
  theory},'' \href{http://dx.doi.org/10.1103/PhysRevD.82.026006}{{\em Phys.
  Rev. D} {\bf 82} (2010)  026006}, \href{http://arxiv.org/abs/0906.4426}{{\tt
  arXiv:0906.4426 [hep-th]}}.

\bibitem{Chesler:2010bi}
P.~M. Chesler and L.~G. Yaffe, ``{Holography and colliding gravitational shock
  waves in asymptotically AdS$_{5}$ spacetime},''
  \href{http://dx.doi.org/10.1103/PhysRevLett.106.021601}{{\em Phys. Rev.
  Lett.} {\bf 106} (2011)  021601}, \href{http://arxiv.org/abs/1011.3562}{{\tt
  arXiv:1011.3562 [hep-th]}}.

\bibitem{Heller:2012km}
M.~P. Heller, D.~Mateos, W.~van~der Schee, and D.~Trancanelli, ``{Strong
  Coupling Isotropization of Non-Abelian Plasmas Simplified},''
  \href{http://dx.doi.org/10.1103/PhysRevLett.108.191601}{{\em Phys. Rev.
  Lett.} {\bf 108} (2012)  191601}, \href{http://arxiv.org/abs/1202.0981}{{\tt
  arXiv:1202.0981 [hep-th]}}.

\bibitem{vanderSchee:2012qj}
W.~van~der Schee, ``{Holographic thermalization with radial flow},''
  \href{http://dx.doi.org/10.1103/PhysRevD.87.061901}{{\em Phys. Rev. D} {\bf
  87} (2013) no.~6, 061901}, \href{http://arxiv.org/abs/1211.2218}{{\tt
  arXiv:1211.2218 [hep-th]}}.

\bibitem{Adams:2013vsa}
A.~Adams, P.~M. Chesler, and H.~Liu, ``{Holographic turbulence},''
  \href{http://dx.doi.org/10.1103/PhysRevLett.112.151602}{{\em Phys. Rev.
  Lett.} {\bf 112} (2014) no.~15, 151602},
  \href{http://arxiv.org/abs/1307.7267}{{\tt arXiv:1307.7267 [hep-th]}}.

\bibitem{Heller:2013oxa}
M.~P. Heller, D.~Mateos, W.~van~der Schee, and M.~Triana, ``{Holographic
  isotropization linearized},''
  \href{http://dx.doi.org/10.1007/JHEP09(2013)026}{{\em JHEP} {\bf 09} (2013)
  026}, \href{http://arxiv.org/abs/1304.5172}{{\tt arXiv:1304.5172 [hep-th]}}.

\bibitem{Casalderrey-Solana:2013aba}
J.~Casalderrey-Solana, M.~P. Heller, D.~Mateos, and W.~van~der Schee, ``{From
  full stopping to transparency in a holographic model of heavy ion
  collisions},'' \href{http://dx.doi.org/10.1103/PhysRevLett.111.181601}{{\em
  Phys. Rev. Lett.} {\bf 111} (2013)  181601},
\href{http://arxiv.org/abs/1305.4919}{{\tt arXiv:1305.4919 [hep-th]}}.

\bibitem{vanderSchee:2013pia}
W.~van~der Schee, P.~Romatschke, and S.~Pratt, ``{Fully Dynamical Simulation of
  Central Nuclear Collisions},''
  \href{http://dx.doi.org/10.1103/PhysRevLett.111.222302}{{\em Phys. Rev.
  Lett.} {\bf 111} (2013) no.~22, 222302},
  \href{http://arxiv.org/abs/1307.2539}{{\tt arXiv:1307.2539 [nucl-th]}}.

\bibitem{Casalderrey-Solana:2013sxa}
J.~Casalderrey-Solana, M.~P. Heller, D.~Mateos, and W.~van~der Schee,
  ``{Longitudinal Coherence in a Holographic Model of Asymmetric Collisions},''
  \href{http://dx.doi.org/10.1103/PhysRevLett.112.221602}{{\em Phys. Rev.
  Lett.} {\bf 112} (2014) no.~22, 221602},
  \href{http://arxiv.org/abs/1312.2956}{{\tt arXiv:1312.2956 [hep-th]}}.

\bibitem{Arnold:2014jva}
P.~Arnold, P.~Romatschke, and W.~van~der Schee, ``{Absence of a local rest
  frame in far from equilibrium quantum matter},''
  \href{http://dx.doi.org/10.1007/JHEP10(2014)110}{{\em JHEP} {\bf 10} (2014)
  110}, \href{http://arxiv.org/abs/1408.2518}{{\tt arXiv:1408.2518 [hep-th]}}.

\bibitem{Chesler:2015wra}
P.~M. Chesler and L.~G. Yaffe, ``{Holography and off-center collisions of
  localized shock waves},''
  \href{http://dx.doi.org/10.1007/JHEP10(2015)070}{{\em JHEP} {\bf 10} (2015)
  070}, \href{http://arxiv.org/abs/1501.04644}{{\tt arXiv:1501.04644
  [hep-th]}}.

\bibitem{Chesler:2015fpa}
P.~M. Chesler, N.~Kilbertus, and W.~van~der Schee, ``{Universal hydrodynamic
  flow in holographic planar shock collisions},''
  \href{http://dx.doi.org/10.1007/JHEP11(2015)135}{{\em JHEP} {\bf 11} (2015)
  135},
\href{http://arxiv.org/abs/1507.02548}{{\tt arXiv:1507.02548 [hep-th]}}.

\bibitem{vanderSchee:2015rta}
W.~van~der Schee and B.~Schenke, ``{Rapidity dependence in holographic heavy
  ion collisions},'' \href{http://dx.doi.org/10.1103/PhysRevC.92.064907}{{\em
  Phys. Rev. C} {\bf 92} (2015) no.~6, 064907},
  \href{http://arxiv.org/abs/1507.08195}{{\tt arXiv:1507.08195 [nucl-th]}}.

\bibitem{Ecker:2021ukv}
C.~Ecker, J.~Erdmenger, and W.~Van Der~Schee, ``{Non-equilibrium steady state
  formation in 3+1 dimensions},'' \href{http://arxiv.org/abs/2103.10435}{{\tt
  arXiv:2103.10435 [hep-th]}}.

\bibitem{Chesler:2013lia}
P.~M. Chesler and L.~G. Yaffe, ``{Numerical solution of gravitational dynamics
  in asymptotically anti-de Sitter spacetimes},''
  \href{http://dx.doi.org/10.1007/JHEP07(2014)086}{{\em JHEP} {\bf 07} (2014)
  086},
\href{http://arxiv.org/abs/1309.1439}{{\tt arXiv:1309.1439 [hep-th]}}.

\bibitem{boyd01}
J.~P. Boyd, {\em {Chebyshev} and {Fourier} Spectral Methods}.
\newblock Dover Books on Mathematics. Dover Publications, Mineola, NY,
  second~ed., 2001.

\bibitem{deHaro:2000vlm}
S.~de~Haro, S.~N. Solodukhin, and K.~Skenderis, ``{Holographic reconstruction
  of space-time and renormalization in the AdS / CFT correspondence},''
  \href{http://dx.doi.org/10.1007/s002200100381}{{\em Commun. Math. Phys.} {\bf
  217} (2001)  595--622},
\href{http://arxiv.org/abs/hep-th/0002230}{{\tt arXiv:hep-th/0002230
  [hep-th]}}.

\bibitem{oldref}
V.~D. Sandberg, ``Tensor spherical harmonics on $s^2$ and $s^3$ as eigenvalue
  problems,'' \href{http://dx.doi.org/10.1063/1.523649}{{\em Journal of
  Mathematical Physics} {\bf 19} (1978) no.~12, 2441--2446},
  \href{http://arxiv.org/abs/https://doi.org/10.1063/1.523649}{{\tt
  https://doi.org/10.1063/1.523649}}.

\bibitem{Athanasiou:2010pv}
C.~Athanasiou, P.~M. Chesler, H.~Liu, D.~Nickel, and K.~Rajagopal,
  ``{Synchrotron radiation in strongly coupled conformal field theories},''
  \href{http://dx.doi.org/10.1103/PhysRevD.81.126001,
  10.1103/PhysRevD.84.069901}{{\em Phys. Rev.} {\bf D81} (2010)  126001},
  \href{http://arxiv.org/abs/1001.3880}{{\tt arXiv:1001.3880 [hep-th]}}.
[Erratum: Phys. Rev.D84,069901(2011)].

\bibitem{Horowitz:2014hja}
G.~T. Horowitz and J.~E. Santos, ``{Geons and the Instability of Anti-de Sitter
  Spacetime},'' \href{http://dx.doi.org/10.4310/SDG.2015.v20.n1.a13}{{\em
  Surveys Diff. Geom.} {\bf 20} (2015)  321--335},
\href{http://arxiv.org/abs/1408.5906}{{\tt arXiv:1408.5906 [gr-qc]}}.

\bibitem{Caron-Huot:2011vtx}
S.~Caron-Huot, P.~M. Chesler, and D.~Teaney, ``{Fluctuation, dissipation, and
  thermalization in non-equilibrium AdS$_5$ black hole geometries},''
  \href{http://dx.doi.org/10.1103/PhysRevD.84.026012}{{\em Phys. Rev. D} {\bf
  84} (2011)  026012}, \href{http://arxiv.org/abs/1102.1073}{{\tt
  arXiv:1102.1073 [hep-th]}}.

\bibitem{Chesler:2011ds}
P.~M. Chesler and D.~Teaney, ``{Dynamical Hawking Radiation and Holographic
  Thermalization},'' \href{http://arxiv.org/abs/1112.6196}{{\tt arXiv:1112.6196
  [hep-th]}}.

\end{thebibliography}\endgroup

\end{document}